# Systematic variation in the apparent burning area of thermonuclear bursts and its implication for neutron star radius measurement

Sudip Bhattacharyya[1]⋆, M. Coleman Miller[2]† and Duncan K. Galloway[3]‡
[1]*Department of Astronomy and Astrophysics, Tata Institute of Fundamental Research, Mumbai 400005, India*
[2]*University of Maryland, Department of Astronomy and Maryland Astronomy Center for Theory and Computation, College Park, Maryland 20742-2421, USA*
[3]*Center for Stellar and Planetary Astrophysics, Monash University, Victoria 3800, Australia*

**ABSTRACT**

Precision measurements of neutron star radii can provide a powerful probe of the properties of cold matter beyond nuclear density. Beginning in the late 1970s it was proposed that the radius could be obtained from the apparent or inferred emitting area during the decay portions of thermonuclear (type I) X-ray bursts. However, this apparent area is generally not constant, preventing reliable measurement of the source radius. Here we report for the first time a correlation between the variation of the inferred area and the burst properties, measured in a sample of almost 900 bursts from 43 sources. We found that the rate of change of the inferred area during decay is anticorrelated with the burst decay duration. A Spearman rank correlation test shows that this relation is significant at the $< 10^{-45}$ level for our entire sample, and at the $7 \times 10^{-37}$ level for the 625 bursts without photospheric radius expansion. This anticorrelation is also highly significant for individual sources exhibiting a wide range of burst durations, such as 4U 1636–536 and Aql X-1. We suggest that variations in the colour factor, which relates the colour temperature resulted from the scattering in the neutron star atmosphere to the effective temperature of the burning layer, may explain the correlation. This in turn implies significant variations in the composition of the atmosphere between bursts with long and short durations.

## 1 INTRODUCTION

Accurate measurement of the radius of a neutron star is a long-standing goal of astrophysics. Such a measurement, along with that of two other independent stellar parameters (e.g. mass and spin frequency), is essential for constraining the theoretically proposed equation of state (EoS) models of neutron star cores (Miller & Lamb 1998; Nath et al. 2002; Bhattacharyya et al. 2005). This constraint provides the only method to probe the cold supra-nuclear core matter, which is a fundamental problem of nuclear physics. Among the three independent neutron star parameters, the measurement of radius is the most difficult one. This is because, whereas the gravitational effect of mass influences the observed motion of the companion star in a binary system, and the observed flux may be modulated by the neutron star spin, the radius exhibits no such effects. As a result, although the mass and the spin frequency of several neutron stars have been measured (Thorsett & Chakrabarty 1999; Lamb & Boutloukos 2008), accurate measurements of the radius are challenging. Thermonuclear bursts provide one of the very few promising methods to measure the neutron star radius.

Thermonuclear (type I) X-ray bursts are observed from neutron star low-mass X-ray binary (LMXB) systems. During a typical burst, the observed X-ray intensity increases rapidly for $\sim 1-5$ s, and then decays slowly for $\sim 10-100$ s. These bursts originate from intermittent unstable nuclear burning of deposited matter on the surfaces of accreting neutron stars (Grindlay et al. 1976; Joss 1977; Lamb & Lamb 1978; Chakrabarty et al. 2003). Observations of bursts can in principle constrain the radii of these stars (Sztajno et al.

⋆ E-mail: sudip@tifr.res.in
† E-mail: miller@astro.umd.edu
‡ E-mail: duncan.galloway@sci.monash.edu.au

1985). This is because the observed spectra of thermonuclear bursts are normally well described with blackbodies (Strohmayer & Bildsten 2006), and hence the inferred radius of the neutron star can be obtained from the relation:

$$R_\infty = (F_\infty/\sigma T_\infty^4)^{1/2} d. \qquad (1)$$

Here $F_\infty$ is the observed bolometric flux, $T_\infty$ is the fitted blackbody temperature, $d$ is the source distance and $\sigma$ is the Stefan-Boltzmann constant.

Unfortunately, current estimates using equation 1 are dominated by systematic uncertainties, including our lack of knowledge about the distance $d$ to sources, their surface gravitational redshift $1+z$, the colour correction factor $f$ due to inelastic Compton scattering and free-free and bound-free emission and absorption in the neutron star atmosphere, and what fraction of the surface is actually emitting at a given time. In addition, as is evident from some of the figures in Galloway et al. (2008) and from our Figure 1, the inferred area of emission can increase, decrease, or go up and down erratically in a given burst. This suggests that we should be cautious in drawing conclusions about the radius from these fits, and motivates us to look for patterns that can reduce the systematic spread.

Here we report the first discovery of a pattern in the inferred emission areas. Using the fits from Galloway et al. (2008), we show that in burst tails the inferred area tends to increase with time for short bursts, but decrease with time for long bursts. There is substantial spread around this trend, suggesting that multiple factors may be at work, but we tentatively suggest that systematic differences in the colour factors related to burst composition (e.g., nearly pure helium for shorter bursts versus mixed hydrogen and helium for longer bursts) may play an important role. In § 2 we describe our analysis and the strength of our correlation. In § 3 we discuss in more detail the possible role of the colour factor, and in § 4 we give our conclusions.

## 2  ANALYSIS AND RESULTS

The large area and good time resolution of the proportional counter array (PCA) detector of the *Rossi X-ray Timing Explorer* (*RXTE*) satellite makes it an ideal instrument for measuring the burst spectral evolution. In order to find a pattern in the $R_\infty$ evolution during the burst decay, we therefore visually examined Fig. 9 of Galloway et al. (2008), which displays the spectral evolution of more than a thousand bursts observed with PCA. As is evident from our Figure 1, $R_\infty$ typically increases with time during burst decay for short duration bursts, whereas it decreases for long duration bursts. Note that in Figure 1 and in our analysis, we have used $R_\infty^2$ (in units of distance squared) instead of $R_\infty$, because $R_\infty^2$ is proportional to the inferred emission area.

A fair quantitative comparison of how the inferred area changes with time requires normalization of both quantities because otherwise, e.g., long bursts automatically have small radius slopes. In addition, photospheric radius expansion (PRE) bursts need to be treated differently from non-PRE bursts because we are interested only in the portion of the burst tail during which the photosphere is at the surface of the star. We therefore adopt the following analysis procedure:

(i) For non-PRE bursts we begin our analysis at the time $t_\text{start}$ of peak inferred bolometric flux, $f_\text{bol,max}$. We end our analysis at the time $t_\text{stop}$ that the inferred bolometric flux drops below a fixed fraction of $f_\text{bol,max}$. Experiments indicated that the precise value of this fraction does not change the results appreciably, hence we present values for a fraction of 0.15. The corresponding fitted squared radii are $R_\text{start}^2$ and $R_\text{stop}^2$.

(ii) For PRE bursts we begin our analysis at the time $t_\text{start}$ at which the emitting region has touched down after expansion, as determined by a local maximum in inferred temperature after the radius maximum. We end our analysis at the time $t_\text{stop}$ when the inferred bolometric flux drops below 0.15 times the bolometric maximum during the entire burst. As before, we define $R_\text{start}^2$ and $R_\text{stop}^2$ corresponding to the start and stop times.

(iii) We normalize the time of the burst so that it runs from 0 to 1 for all bursts: $t \to (t - t_\text{start})/(t_\text{stop} - t_\text{start})$.

(iv) We define $R_\text{av}^2 \equiv (R_\text{start}^2 + R_\text{stop}^2)/2$. Using this, we define a normalized squared radius: $R^2 \to (R^2 - R_\text{start}^2)/R_\text{av}^2$.

(v) The trend in radius-squared with time is then the best-fit linear slope $R_\text{sl}^2$, which can be considered as an average $dR^2/dt$.

(vi) In a similar fashion, we fit a temperature slope $T_\text{sl}$ to the data, where we have normalized the fitted temperature $T$ to the start and end temperatures. Our definitions imply that both $R_\text{sl}^2$ and $T_\text{sl}$ are constrained to lie in the range [-2,+2].

As defined, $R_\text{sl}^2$ and $T_\text{sl}$ are independent of quantities such as $d$ and $1+z$ that are fixed for a given source and independent of any other physical parameters that do not evolve during a burst. Our correlation is therefore between the burst duration $t_D \equiv t_\text{stop} - t_\text{start}$ (or actually the $\log_{10}$ of the duration in seconds) and either $R_\text{sl}^2$ or $T_\text{sl}$.

We include in our analysis only those bursts with at least $n_\text{qp}$ (= 6 in our case) qualifying measurements between $t_\text{start}$ and $t_\text{stop}$, as determined from the data tables of Galloway et al. (2008), to avoid spuriously large error bars on the slopes. However, note that a substantial change in $n_\text{qp}$ value does not change our result. In addition, we exclude bursts from Galloway et al. (2008) for which a thermonuclear origin is not certain, specifically the bursts from the Rapid Burster, 4U 1746-37 and Cyg X-2. However, any cut on the reduced $\chi^2$ corresponding to the blackbody fitting of the burst spectra does not change our result, and we do not apply such cuts. This leaves 877 bursts from 43 sources, of which 252 are PRE bursts and 625 are not PRE bursts.

In Figure 2, we plot $R_\text{sl}^2$ and $T_\text{sl}$ against $t_D$ for all 877 bursts. Figure 2 shows that $R_\text{sl}^2$ is negatively correlated and $T_\text{sl}$ is positively correlated with $t_D$. A Spearman rank linear correlation test shows that the $R_\text{sl}^2$ correlation is significant at better than the $10^{-45}$ level, and the $T_\text{sl}$ correlation is significant at the $1.5 \times 10^{-38}$ level. These are the first such correlations that have been demonstrated between blackbody fit parameters and the duration of the bursts.

As Figure 3 shows, the correlation also holds for individual sources, so this is not simply a statement about the population. Here we display $R_\text{sl}^2$ versus $t_D$ for two sources, 4U 1636–536 and Aql X-1, that have a broad range of burst durations. The significance of the correlation as determined by a Spearman rank test is $5 \times 10^{-14}$ for 4U 1636–536, and $2 \times 10^{-7}$ for Aql X-1.

Since PRE and non-PRE bursts are different in some physical aspects, we tested whether the $R_{\rm sl}^2$ vs. $t_{\rm D}$ correlation holds for each set separately. Figure 4 demonstrates that this correlation holds for just the 625 non-PRE bursts, at a significance level of $7 \times 10^{-37}$. However, the 252 PRE bursts by themselves have a correlation in the other direction (longer bursts have larger $R_{\rm sl}^2$), at a significance level of $4 \times 10^{-7}$. The PRE bursts are concentrated at low durations, and form a "cap" to the overall distribution that actually increases the overall significance of the negative correlation. It could therefore be that in this narrow duration range the spread of $R_{\rm sl}^2$ values is the dominant effect, or it could be that somewhat different processes are operating than for the non-PRE bursts. Further research is needed.

## 3 DISCUSSION

We now discuss the origin of the observed $R_{\rm sl}^2$ vs. $t_{\rm D}$ correlation (which, due to our definitions, has a roughly one to one relation with the $T_{\rm sl}$ vs. $t_{\rm D}$ relation). Why should $R_\infty^2$ evolve during the burst decay at all? Van Paradijs & Lewin (1986) discussed that $R_\infty^2$ tends to zero as the blackbody flux from the burst tends to zero towards the very end of the burst tail. However, this does not affect the observed $R_{\rm sl}^2$ vs. $t_{\rm D}$ correlation, as this correlation exists and remains extremely strong for a wide range of $f_{\rm mf}$ (0.05, 0.15, 0.25), where $f_{\rm mf}$ is the fraction of the maximum bolometric burst flux corresponding to $t_{\rm stop}$ (see § 2). Therefore, in order to understand the $R_\infty^2$ evolution and the correlation, we note that as a result of gravitational redshift and spectral hardening, the actual temperature ($T_{\rm BB}$) and radius ($R_{\rm BB}$) as measured at the stellar surface are related to the inferred values by the following relations (Sztajno et al. 1985):

$$T_{\rm BB} = T_\infty (1+z)/f; \tag{2}$$

$$R_{\rm BB} = R_\infty f^2/(1+z). \tag{3}$$

Here $f$ is the spectral hardening factor (London et al. 1986; Madej et al. 2004), which accounts for hardening due to the scattering of photons by the electrons in a neutron star atmosphere. From equation 3 we see that for a given source with fixed $z$, $R_\infty^2$ can change only if the colour factor $f$ changes, and/or the actual burning region area ($\propto R_{\rm BB}^2$) evolves, perhaps in combination with the change in the plausible anisotropy of X-ray emission. The colour factor $f$ is a function of the chemical composition of the neutron star atmosphere, the actual surface temperature $T_{\rm BB}$ and the stellar surface gravity (Madej et al. 2004; Majczyna et al. 2005). Since during the burst decay $T_{\rm BB}$ changes (primarily decreases) and the atmospheric chemical composition may evolve because of ongoing nuclear reactions and/or fluid dynamical processes, $f$ is expected to change. It is not as clear why $R_{\rm BB}^2$ should change in the required fashion, but we will now examine both options. We note that there is a significant spread in the correlation, thus even if one effect dominates it could be that the other is the primary cause of the dispersion in the relation.

First, suppose that $R_{\rm BB}^2$ is constant throughout the decay of all bursts from a given source, but $f$ changes. Then if $R_\infty^2$ increases with time, $f$ has to decrease with time, and vice versa. This, as well as the natural expectation that $T_{\rm BB}$ primarily decreases during decay, imply that $f$ decreases as $T_{\rm BB}$ decreases for short duration bursts, and $f$ increases as $T_{\rm BB}$ decreases for long duration bursts. According to the current theoretical results, the former $f$ vs. $T_{\rm BB}$ relation is possible for lower iron abundance, whereas the latter relation is possible for higher iron abundance (Madej et al. 2004; Majczyna et al. 2005). For example, for an atmosphere with a surface gravity of $g = 10^{14.5}$ cm s$^{-2}$ and a composition of $N_{\rm He}/N_{\rm H} = 0.11$ with no heavier elements, Madej et al. (2004) find that $f$ increases monotonically with $T$, from $f = 1.33$ at $T = 10^7$ K to $f = 1.73$ at $T = 2.5 \times 10^7$ K. In contrast, for the same surface gravity and He/H ratio but with an additional supersolar iron abundance of $N_{\rm Fe}/N_{\rm H} = 10^{-3}$, Majczyna et al. (2005) find that $f$ decreases from 1.49 at $T = 10^7$ K to 1.33 at $T = 2 \times 10^7$ K, only to rise again to 1.43 at $T = 2.5 \times 10^7$ K.

Majczyna et al. (2005) attribute some of the differences to the greater importance of thermal absorption in atmospheres with more iron, but the complexity of the trends suggests that much further study will be required. If compositional differences are the main cause of the differing radius slopes, it could imply that neutron star atmospheres during decay are less metal rich for short duration bursts, and more metal rich for long duration bursts. This could be because short bursts are primarily He-rich (and hence burn by the $3\alpha$ process) and long bursts are mixed H-He bursts (and are thus dominated by rapid proton captures). Short and long bursts therefore have different energetics, ashes, and fluid dynamics (Strohmayer & Bildsten 2006). As an example, convective overturn could bring the ashes of burning to the photosphere (Joss 1977; Weinberg et al. 2006). However, we note that this is only a possibility, and detailed theoretical work, especially to find out if the burning products can be pushed up into the atmosphere, is essential.

More quantitatively, we can ask whether the theoretically calculated range of $f$ can explain the observational range of $R_{\rm sl}^2$. If $R_{\rm BB}^2$ does not change during the burst decay, then from our equations we find $R_{\rm sl}^2 = 2(1 - f_{\rm ratio}^4)/(1 + f_{\rm ratio}^4)$. Here, $f_{\rm ratio} = f_2/f_1$, where $f_1$ and $f_2$ are the colour factors corresponding to the first point and the last point of the burst decay duration respectively. For the chemical compositions assumed in Madej et al. (2004) and Majczyna et al. (2005), the theoretical value of $f_{\rm ratio}$ is always within the range of 0.67 to 1.5. The $f_{\rm ratio}$ range of $0.67 - 1.5$ implies a range of $-1.34$ to $1.34$ for $R_{\rm sl}^2$. Fig. 2 shows that most of the empirical $R_{\rm sl}^2$ values are within this range, and the rest are consistent with the lateral spread (or dispersion) of the $R_{\rm sl}^2$ vs. $t_{\rm D}$ correlation. This is consistent with the assumption that the change of $f$ can explain the main trend in the evolution of $R_\infty^2$.

Now suppose that $R_{\rm BB}^2$ evolves during the burst decay, but $f$ does not. Given that the burning area must be less than or equal to the surface area of the star, indeed much less in many cases if $f$ is constant, then if the burning region is contiguous (e.g., a single spot) we would expect burst oscillations in most cases (if the burning were not axisymmetric with respect to the neutron star spin axis; Chakrabarty et al. 2003; Bhattacharyya et al. 2005; Strohmayer & Bildsten 2006). However, for most of the bursts with $R_\infty^2$ evolution (and hence $R_{\rm BB}^2$ evolution with the current assumption) dur-ing decay, we do not find burst decay oscillations (Galloway et al. 2008). Moreover, if $R_{\rm BB}^2$ evolution gives rise to the strong $R_{\rm sl}^2$ vs. $t_{\rm D}$ correlation, then we expect that the burst oscillation fractional amplitude should decrease with time during decay for short duration bursts, while it should increase for long duration bursts. We do not find such a trend clearly from nine bursts with published oscillation amplitude evolution (Miller 2000; Muno et al. 2002; Galloway et al. 2008).

It is possible to evade this prediction if the burning is not contiguous, but is rather distributed as a large number of independent pools of fuel scattered over the surface, perhaps confined by small-scale but strong magnetic fields. Such a distribution would lead to very low oscillation amplitudes that would plausibly be undetectable, and one might imagine that the pool sizes could shrink or grow depending on circumstances. This is, however, an ad hoc picture that is difficult to evaluate, whereas the colour factor model has, at the lowest order, quantitative consistency with the basic observations.

## 4   CONCLUSIONS

In this Letter, we report the first discovery of a pattern in the seemingly erratic $R_\infty^2$ evolution during the burst decay: for short bursts, $R_\infty^2$ tends to increase with time, whereas for long bursts $R_\infty^2$ tends to decrease with time. If this pattern can be understood theoretically, it will help reduce the currently significant uncertainties in radius estimates for neutron stars from burst modelling. We propose that systematic changes in the colour factor likely drive the main trend, but that changes in the actual burning area may contribute significantly to the spread in the relation (and possibly somewhat to the trend as well).

Whatever evolution is eventually shown to cause the $R_{\rm sl}^2$ vs. $t_{\rm D}$ correlation, it will have a strong impact on our understanding of thermonuclear bursts. For example, if $R_{\rm BB}^2$ evolution is the main driver of the correlation (or even of the spread in the correlation instead of the main trend), it will be a challenge to explain (1) why the actual burning area changes during the burst decay and (2) why this change and the burst decay duration are related. On the other hand, if $f$ evolution is the primary cause of the observed correlation, it will be important to produce detailed models that match this evolution by, e.g., changes in the chemical composition of the photosphere.

Finally, we emphasize that if future studies establish the true cause of the correlation and the lateral spread, it may improve the precision and accuracy of neutron star radius estimates using $R_\infty^2$. This will require detailed theoretical modelling, as well as observations with upcoming satellites such as *Astrosat*.

## ACKNOWLEDGMENTS

We thank Fred Lamb for useful discussions, and also thank the referee for a helpful report that clarified this manuscript. This work was supported in part by US NSF grant AST 0708424.

## REFERENCES

Bhattacharyya S., Strohmayer T. E., Miller M. C., Markwardt C. B., 2005, ApJ, 619, 483
Chakrabarty D., Morgan E. H., Muno M. P., Galloway D. K., Wijnands R., van der Klis M., Markwardt C. B., 2003, Nature, 424, 42
Galloway D. K., Muno M. P., Hartman J. M., Psaltis D., Chakrabarty D., 2008, ApJSS, 179, 360
Grindlay J., Gursky H., Schnopper H., Parsignault D. R., Heise J., Brinkman A. C., Schrijver, J., 1976, ApJ, 205, L127
Joss P. C., 1977, Nature, 270, 310
Lamb D. Q., Lamb F. K., 1978, ApJ, 220, 291
Lamb F. K., Boutloukos, S., 2008, ASS L352, 87
London R. A., Taam R. E., Howard W. M., 1986, ApJ, 306, 170
Madej J., Joss P. C., Róaska, A., 2004, ApJ, 602, 904
Majczyna A., Madej J., Joss P. C., Róaska, A., 2005, A&A, 430, 643
Miller M. C., 2000, ApJ, 531, 458
Miller M. C., Lamb F. K., 1998, ApJ, 499, L37
Muno M. P., Özel F., Chakrabarty D., 2002, ApJ, 581, 550
Nath N. R., Strohmayer T. E., Swank J. H., 2002, ApJ, 564, 353
Strohmayer T. E., Bildsten, L., 2006, in Compact Stellar X-Ray Sources, eds. Lewin W. H. G., van der Klis M., Cambridge Univ. Press, 113
Sztajno M., van Paradijs J., Lewin W. H. G., Trumper J., Stollman G., Pietsch W., van der Klis M., 1985, ApJ, 299, 487
Thorsett S. E., Chakrabarty D., 1999, ApJ, 512, 288
van Paradijs J., Lewin H. G., 1986, A&A, 157, L10
Weinberg N. N., Bildsten, L., Schatz, H., 2006, ApJ, 639, 1018

This paper has been typeset from a T$_{\rm E}$X/ L$^{\rm A}$T$_{\rm E}$X file prepared by the author.

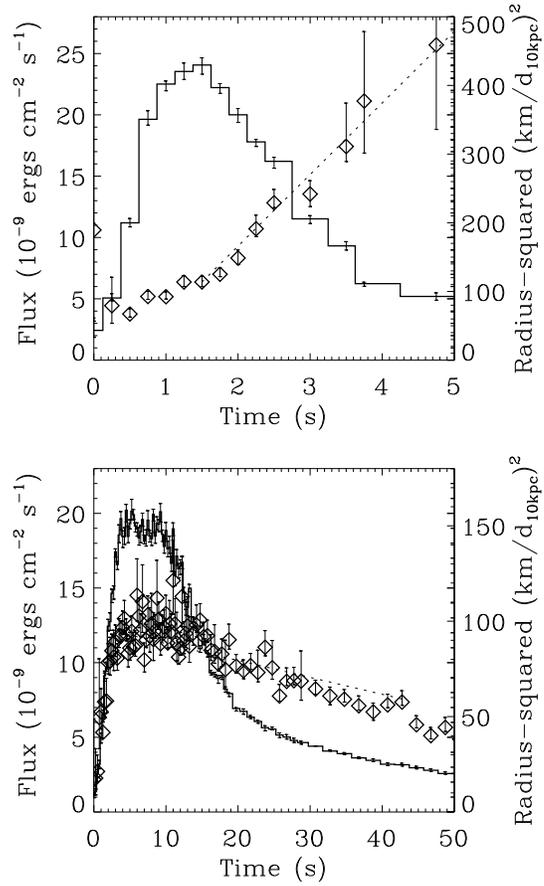

**Figure 1.** *RXTE* PCA data of two thermonuclear X-ray bursts from the neutron star LMXB 4U 1636-536 (upper panel: burst #8; lower panel: burst #166; Galloway et al. 2008). The histograms show the bolometric flux profiles and the diamonds exhibit the spectral blackbody radius-squared profiles. The dotted lines correspond to the best-fit radius-squared slopes (see § 2). A one standard deviation error bar is attached to each data point. The distance of the source is $d_{10\rm kpc}$ in units of 10 kpc. The radius-squared slope (see § 2 for definition) is 1.52 for the upper panel, and $-0.45$ for the lower panel. This figure shows that the fitted blackbody radius-squared typically increases during the burst decay for short bursts, and decreases for long bursts (see Fig. 2 to find this behaviour for more bursts).

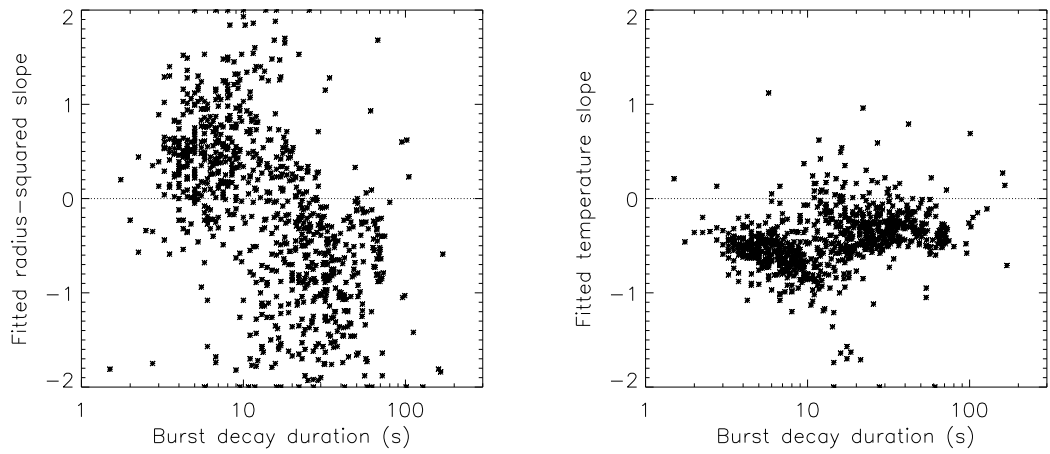

**Figure 2.** *Left panel*: Slope of blackbody radius-squared during burst decay vs. burst decay duration. Each point corresponds to one burst, and there are 877 bursts from 43 sources shown in the panel. This panel indicates a strong negative correlation between the observed radius-squared slope and the decay duration (see text). Note that due to our definition of the radius-squared slope, if the initial radius or final radius is very small compared to intermediate values, the slopes tend to the extremal values of +2 or -2. *Right panel*: Similar to the left panel, but for the blackbody temperature slope instead of the radius-squared slope.

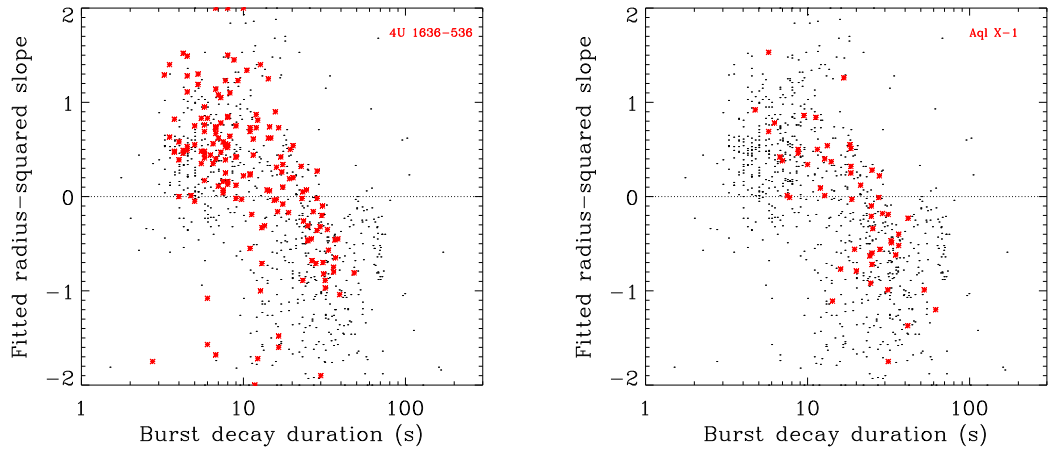

**Figure 3.** Slope of blackbody radius-squared during burst decay vs. burst decay duration. The 877 light black points of each panel are for all the bursts shown in the left panel of Fig. 2. *Left panel*: red points are for 159 bursts from the source 4U 1636-536. *Right panel*: red points are for 54 bursts from the source Aql X-1. These panels indicate a negative correlation between the observed radius-squared slope and the decay duration even for these two individual sources (see text).

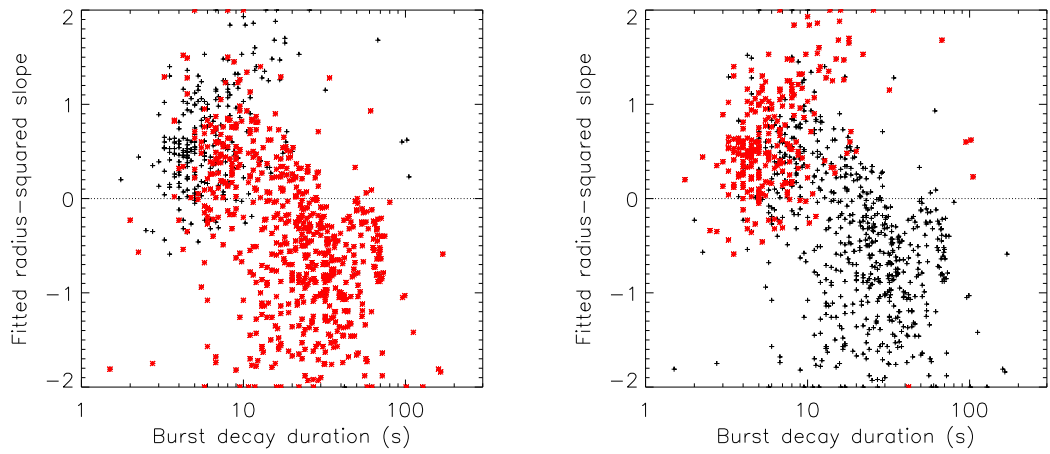

**Figure 4.** Slope of blackbody radius-squared during burst decay vs. burst decay duration. The 877 black and red points of each panel are for both PRE and non-PRE bursts shown in the left panel of Fig. 2. *Left panel*: red points are for 625 non-PRE bursts. This panel indicates a strong negative correlation between the observed radius-squared slope and the decay duration even for only non-PRE bursts (see text). *Right panel*: red points are for 252 PRE bursts. This panel indicates a weak positive correlation between the observed radius-squared slope and the decay duration for PRE bursts (but see text).